\documentstyle[prl,psfig,preprint,aps]{revtex}
\begin{document}
\draft
\title{Detection of Nonlinear Coupling and \\
its Application to Cardiorespiratory Interaction}

\author{Guillermo J. Ortega and Diego A. Golombek}
\address{Centro de Estudios e Investigaciones, Universidad Nacional
de Quilmes, \\
R.S. Pe\~na 180, (1876) Bernal, Provincia de Buenos Aires, Argentina}
\date{\today}
\maketitle
\begin{abstract}
{We present here a modification of the Lagrangian measures technique,
which allows a reliable detection of interdependency among simultaneous
measurements of different variables. This method is applied to a 
simulated multivariate
time series and to a bivariate cardiorespiratory signal. By using
this methodology, it is possible to reveal a nonlinear interaction
among cardiac and respiration rhythms in pathological conditions.
}

\end{abstract}
\pacs{PACS numbers: 87.22.As, 05.45.+b}

The study of natural complex phenomena has urged
for the reformulation of signal processing for
nonlinear systems. Moreover, with the advent of
chaos theory and the "suspicion" that many natural phenomena
may behave in a chaotic, although deterministic,
way, new analytical methods have been developed.
Many of these complex behavior
arise as the interaction of the variables involved
in the process under consideration. This is especially true
in the case of physiological monitoring, where the 
signals are correlated with one another by feedback mechanisms.
Very recently \cite{SR98} \cite{RS99}, cardiorespiratory interaction 
has attracted the interest of the nonlinear community as a subject
area where application of new methodologies can be tested.

Cardiorespiratory interaction is one of these examples
where traditional tools have been used in
order to gain information about its underlying dynamics.
In this case, an interdependency between heart rate (HR) and respiration
(R) rhythms, in physiological conditions, is almost present
and is known as Respiratory Sinus Arrhithmia (see \cite{AG81}
and references therein).
This phenomenon is especially evident in the usual analysis
by means of linear tools, like power spectrum. Accordingly,
the power spectrum of the heart rate contains a peak centered
at the respiratory frequency  \cite{AG81} \cite{RG94}. However, 
there may exist situations where this peak will not be present. 
This fact is interpreted as a blockade in the interaction
between both systems.

In this Letter, we use a recently developed technique
\cite{O95,O96}, originally aimed to detect hidden
frequencies in time series, with the purpose to demonstrate the 
interdependence among cardiac and respiratory signals, even
in pathological conditions, where traditional tools,
like power spectrum, show no signs of interactions.
As we will explain below, a simple modification of 
our original approach will allow us to deal with this
rather noisy and short time series. 
For illustration, we will show the technique in a
well known system, the Lorenz system, in
order to understand the basic steps.

From the point of view of nonlinear dynamics \cite{AB93,ER85,SY91}, 
a time series is considered as the "output" or observable
of a dynamical system, which can be described by the $p$-first 
order differential equation.
In the case of chaotic behavior, the system must be necessarily 
multivariate (that is, $p > 2$). 
Embedding techniques \cite{SY91} provide a way to reconstruct geometric 
(attractor structure) and statistical information of the original system 
by constructing an equivalent representation of it, using time-delayed 
coordinates. Given a time series of $m$ values, an $n$-dimensional vector
can be constructed:
\begin{eqnarray*}
\bar {x_i}  = (x_i, x_{i+ \tau},...,x_{i+(n-1).\tau})
\end{eqnarray*}
where $\tau$ is the time delay. In this way, from a single set 
of observations, multivariate vectors in the reconstructed
$n$-dimensional space are used to trace the orbit of the system.

One of the many ways to describe the dynamics is by 
studying the long-term behavior of the system.
When performing a statistical description of dynamical systems,
a central role is played by the natural (or {\it physical}) probability
measure, which describes where the orbit has been, and whose
integral over a volume of state space counts the number of points 
within that volume \cite{AB93,ER85}. By performing a partition in the
reconstructed phase space of dimension $n$, this probability density can 
be estimated as,
\begin{equation}
\hat{\mu} ({\bf x}_i) = \frac{n({\bf x}_i)} {\sum_{j} n({\bf x}_j)}
\label{eq1}
\end{equation}
with $n({\bf x}_i)$ equal to the number of points in partition 
element $i$ around the point ${\bf x}_i$.
Using this approach, we have recently  proposed a new 
technique to detect hidden frequencies in chaotic time series 
\cite{O95,O96}. Here we use an extension of this technique
which gives a reliable detection of intrinsic frequencies in
noisy and short time series.

Basically, the procedure is the following: for a given time 
series, a fixed, relatively high embedding 
dimension is chosen (high enough in order to unfold the 
geometrical structure of the attractor), and  
a reconstruction for a given range of time lags $\tau$ 
(the number of lags should be several times the 
correlation length of the time series) is then performed. 
For each $\tau$ the density of points the 
trajectory encounters as it evolves is calculated. 
In this case, the density along the reconstructed 
trajectory is estimated as 
\begin{equation}
\hat{\mu} ({\bf x}_i,\tau) = \frac{n({\bf x}_i,\tau)} 
{\sum_{j} n({\bf x}_j,\tau)}
\label{eq2}
\end{equation}
that is, every ${\bf x}_i$ is ordered consecutively along the 
trajectory. Roughly speaking, this is an estimate 
of the probability measure the system trajectory 
encounters as it evolves. In order to estimate the 
density given by Eq. \ref{eq1} we have used small spheres 
around each point ${\bf x}_i$, and count the 
number of other points ${\bf x}_j$ inside this volume. 
Typically, we have used a radius of 5\% to 20\% of the 
total extent of the attractor. In this way, a new time 
series may be constructed with the density data, 
which now gives information of the different regions in 
the reconstructed phase space that the 
system visits. This information reveals recurring motion in 
the phase space, which is ultimately 
transferred to the observable time series as periodicity 
information. A periodogram \cite{PF88} \cite{OS89} is 
performed over this 
density time series, $\hat{P}$, for each $\tau$ (see 
reference \cite{PF88} for numerical implementation) as

\begin{equation}
\hat{P}(f_k, \tau ) = {1 \over N^2} \left[ \sum_{j=0}^{N-1}
\hat{\mu} ({\bf x}_i,\tau) e^{{i 2 \pi j k} \over N} \right]^2
\label{eq3}
\end{equation}
with $N$ representing  the number of data points in 
the density time series.

In order to plot all the periodograms in a single 
graph we have used a gray-scale map, and 3 contour
curves were superimposed in each plot in order to
clarify visual inspection.  

A note about the use of the periodogram is in order here. 
The periodogram is based in the direct Fourier transformation
of the signal \cite{OS89}. If the signal comes from a 
deterministic system, no further modification is needed in
order to the correct interpretation of the periodogram in
terms of the Fourier transform. However, in the case of
more noise-like signals, it is best to introduce statistical
analysis, because for each frequency $f_k$, $P(f_k)$ is
in fact a random variable, which can introduce fluctuations
in the estimation process. This is usually accomplished
by using {\it averaging} \cite{OS89}, with several realizations
of the estimate. In our case, we have used plain periodograms
because in plotting them side by side, several realizations 
of the periodograms is in fact equivalent to the process of
averaging. True frequencies must remain constant in the
embedding process, so any kind of fluctuations in the
estimate can be readily detected. 

The Lorenz system \cite{L63} is a model proposed to explain 
the convective dynamics in the atmosphere (known as 
the Rayleigh-Benard convection), with the 
following variables:
\begin{eqnarray}
{\dot x} & = & -s x + s y  \nonumber \\
{\dot y} & = &  -y + rx - xz  \\
\label{eq4}
\nonumber
{\dot z} & = & -bz + xy  \nonumber
\end{eqnarray}
and the standard parameters $s$ = 10.0, $r$ = 28.0 and 
$b$ = 2.66, that yields a chaotic regime. The $x$ 
coordinate is proportional to the velocity of the 
circulating flux, while the $z$ coordinate represents 
the distortion of the temperature with respect to a linear 
profile between the upper and lower 
temperature. Figures 1(a) and 1(b) (and first and third 
stripes in Figure 1(c)) shows the power spectra 
of the $x$ and $z$ coordinates, which look totally different, 
forcing the preliminary conclusion that there 
is no dynamical connection between both phenomena. Moreover, 
the power spectrum of the $x$ 
coordinate shows no characteristic frequency at all. 
Its power is mainly distributed in the lower 
band; in contrast, the power spectrum of the $z$ coordinate 
shows a very definite frequency around 
harmonic number 54. However, both time series belong to 
the same system. It must be remarked that more involved linear 
methods, like cross-correlation and coherence gives no further 
insight in this problem. The above fact reveals the
main disadvantage in applying spectral methods for analyzing 
time series coming from chaotic 
systems. We have used the $x$ variable as our experimental
time series with 2048 data points. An 
embedding dimension of 3 has been used, and a 10\% radius of the 
attractor was employed in order 
to estimate the density. The main panel of Figure 1(c) shows the 
power spectrum of the density 
time series after applying our method to the $x$ coordinate, 
for each $\tau$ of the embedding process. As 
the $\tau$ parameter varies, the strong frequency which appears 
in the $z$ coordinate is remarkably recovered, as well as some
of the frequencies in the lower band. 
Inversely, it is also possible to use the $z$ coordinate as
the experimental one. Figure 2 shows the same as Figure 1
this time using the $z$ coordinate. Although the power spectrum of 
this coordinate shows a single sharp peak, perhaps 
superimposed to a continuous background, our method allows
to predict the existence of lower band frequencies in its
dynamics. This is evident in the power spectrum of the 
$x$ or $y$ coordinates.

Two additional points are 
noteworthy. Some spurious frequencies may appear in this 
procedure, for example due to an effect 
of the finite size of time series. However, these are easily 
detected because they do not remain 
constant along the $\tau$ scan. By varying the $\tau$ parameter in the 
reconstruction, a nonlinear 
transformation is performed, and intrinsic properties of the 
system must remain constant along this
procedure, a fact guaranteed by the embedding theorems \cite{SY91}
In fact, we are seeking that $\hat{P}(f_i, \tau) = 
\hat{P}(f_i)$, independently of the particular $\tau$ used. 
The second point is that intrinsic frequencies (as is the
case of harmonic number 54) may disappear for some values of the 
$\tau$ parameter. This is due to the 
embedding itself, which is only guaranteed to give faithful 
results in the case of infinitely long data 
sets and without noise. Some values of the time lag $\tau$ may 
introduce false information, or even hide 
important frequencies. By performing a scan along a series of $\tau$
we can be assured of extracting 
information of the system itself.

Now, we use our approach in the multivariate 
time series from heart rate and respiration rhythms. 
The experiment is fully described in reference \cite{RG94}. 
This is a multivariate time series recorded 
from a patient in the sleep laboratory of the Beth Israel 
Hospital in Boston. The HR-record is the heart 
rate, the R-record is the chest volume (respiration force) 
and the BOC-record is the blood oxygen 
concentration (measured by ear oximetry). 2048 data 
points were chosen from the total record, 
corresponding to the periods of the sleep apnea episodes 
of the patient. Figure 3 shows the results 
of applying our analysis to the respiration force time 
series. We have used an embedding dimension of 4 for
the reconstruction process and a radius of the ball
of 15\% of the reconstructed attractor. 

Classical power spectrum analysis of 
the HR-record shows a strong peak at harmonic number 
22 (which corresponds to a frequency of 
0.021 Hz). However, the R-record shows several peaks around 
the harmonic number 260 (0.254 
Hz), and no traces of the main frequency of the HR-record.
Therefore, it could be concluded that both 
variables are (linearly) independent. In the original
analysis \cite{RG94}, based on standard spectral 
analysis there is no mention at all of any kind of
coupling among both variables in this pathological
case, i.e. sleep apnea, contrary to the case of
the RSA in the physiological case.  

By applying our analysis to the R-record 
(Figure 3(c), main panel), 
it appears that the main frequencies in the 
reconstructed system are those related with the HR record, 
and some traces of the original R time series
can be observed (around $\tau = 25$). 
This fact, as in the previous case of the
Lorenz system, could be considered as a justification of 
a nonlinear relation between both variables, and, moreover, 
suggests that they are related with the same dynamics. 
However, in this particular case, the flow of information
is in the oposite direction as in the case of the RSA.

As in the case of the Lorenz system, we can apply the
procedure to the other variable, that is, the HR record
(Figure 4). Although some traces of the
R record are recovered (specially around $\tau$ = 90)
almost all of the power is concentrated in the characteristic
frequency of the HR. This would imply a very weak coupling
from the heart rate to the respiration variable. In the
case of the Lorenz system, the equations shows that both
variables used, $x$ and $z$ are in fact coupled to each
other. That is why it is possible to recover the whole
spectrum by using whatever of both variables.
One can conclude from the above argument that the
coupling among HR and R variables could have a preferential
flow direction from the HR to the R, in this pathological 
condition.

Experience in numerical models shows that hidden
frequencies, as detected by our approach, are caused
fundamentally by chaotic systems, when one of the variables
"fails" in projecting its periodic content over the measured
time series.
In a linear system, or a nonlinear system without a
chaotic behavior, this does not happen. Considering
the cardiorrespiratory interaction, the modulation of 
HR by  respiration is clearly seen 
in the power spectrum of the HR signal under physiological conditions.
However, in
the pathological condition we are considering, this
coupling is not revealed by linear analysis, and moreover, it
appears to be in the opposite direction. It could be argued that the
underlying dynamics has changed from
a linear/nonlinear state to a chaotic behavior.
This could be a reason for the dissapearance of the
RSA phenomena and the appearance of an inverse coupling. This
fact could also be in accordance with
the so-called "dynamical diseases" \cite{MG88} where
an abrupt change in the dynamics of the
interacting variables is evidenced.

It is very worthy to mention that we have used the
data set as it is. No further filter or smoothing routines
have been applied to the data, beyond those of the
recording process and to the conversion from the ECG to Heart
rate data \cite{RG94}. We think that it is
very important, from the point of view of nonlinear
dynamics analysis, to minimize the use of any kind of 
"treatment" of the experimental record, which may
alter the underlying dynamics, especially in this
short and very noisy data sets.

Data requirements for applying this method are almost
the same as in any classical spectral analysis.
Stationarity \cite{P91} of the data is the most
important prerequisite, because of the probability
estimate that is being calculated. The method is fairly
robust against noise contamination, mainly
because of the density estimation step, as we have shown
with other methods \cite{O96,OL98}. Other than
these considerations, the algorithm is of wide use,
and could be applied to any kind of experimental
data, from biological to geophysical and so on.
We are currently applying this analysis to
physiological data derived from biological rhythmicity
recording in mammals \cite{OG94}.
It should be noted that even more information could be
gathered from the reconstructed density
time series by applying nonlinear analytical methods
\cite{OL98}. However, the use of power spectra is
still the most common tool in time series analysis because
of its simplicity and ease of
interpretation \cite{PF88,B90}, and here we propose to extend its
well known capabilities with the aids of
nonlinear methods. The most important benefits derived
from our methodology might be observed
in the case of several variables, where one or more
of them are not available experimentally. In this
way, more information could be achieved that the one
currently available from classical spectral
analysis techniques.

The authors are members of CONICET Argentina.
Studies in D.A.G.'s laboratory are funded by
grants from CONICET, UNQ, ANPCyT and Fundaci\'on Antorchas.

\begin{figure}[tbh]
\begin{center}
\
\psfig{file=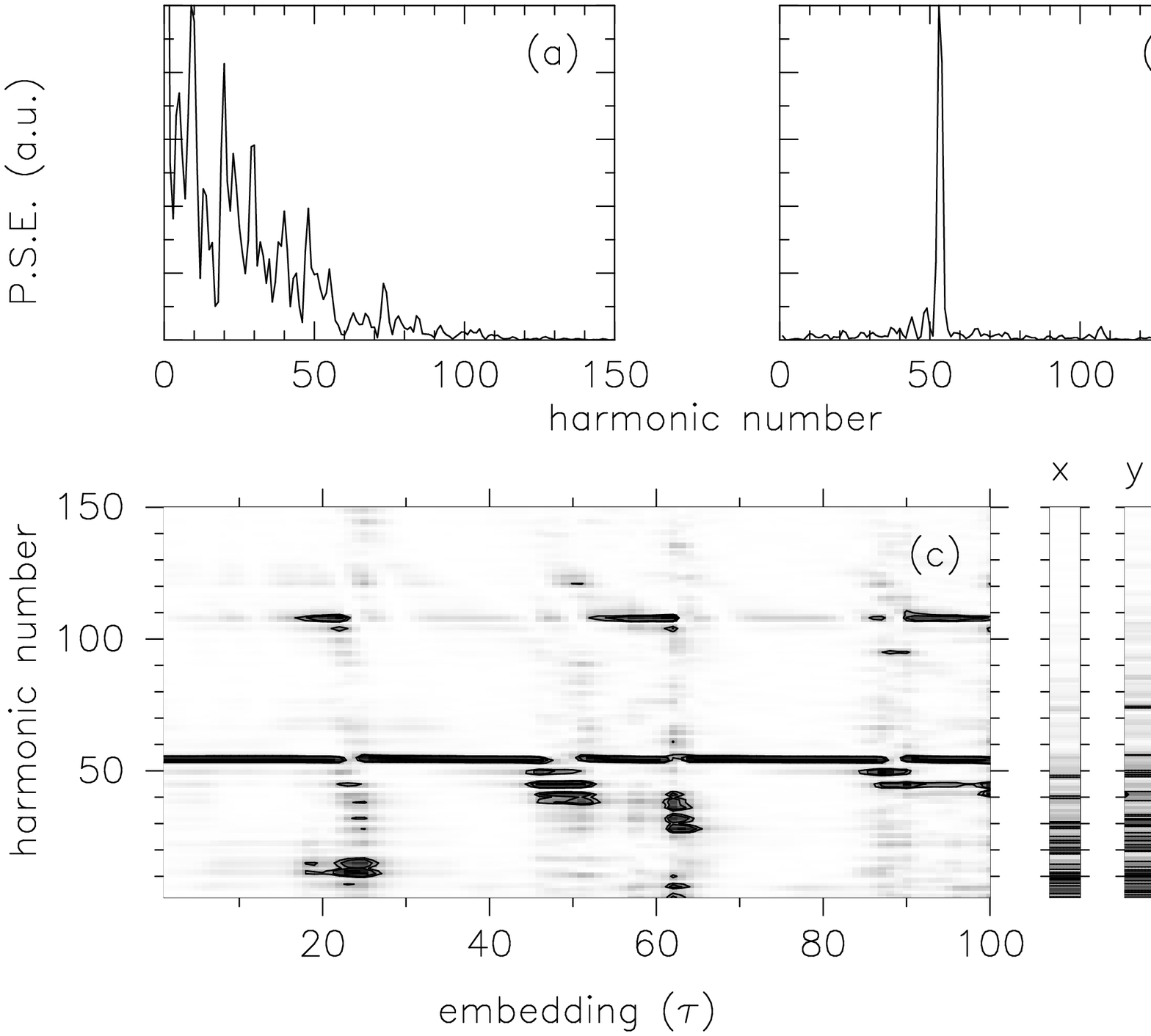,height=4.0in}
\caption{Spectral analysis and nonlinear coupling in the
Lorenz system. Reconstruction using the $x$ coordinate.}
\end{center}
\end{figure}

\newpage

\begin{figure}[tbh]
\begin{center}
\
\psfig{file=Fig2_ps.ps,height=4.0in}
\caption{Spectral analysis and nonlinear coupling in the
Lorenz system. Reconstruction using the $z$ coordinate.}
\end{center}
\end{figure}

\newpage

\begin{figure}[tbh]
\begin{center}
\
\psfig{file=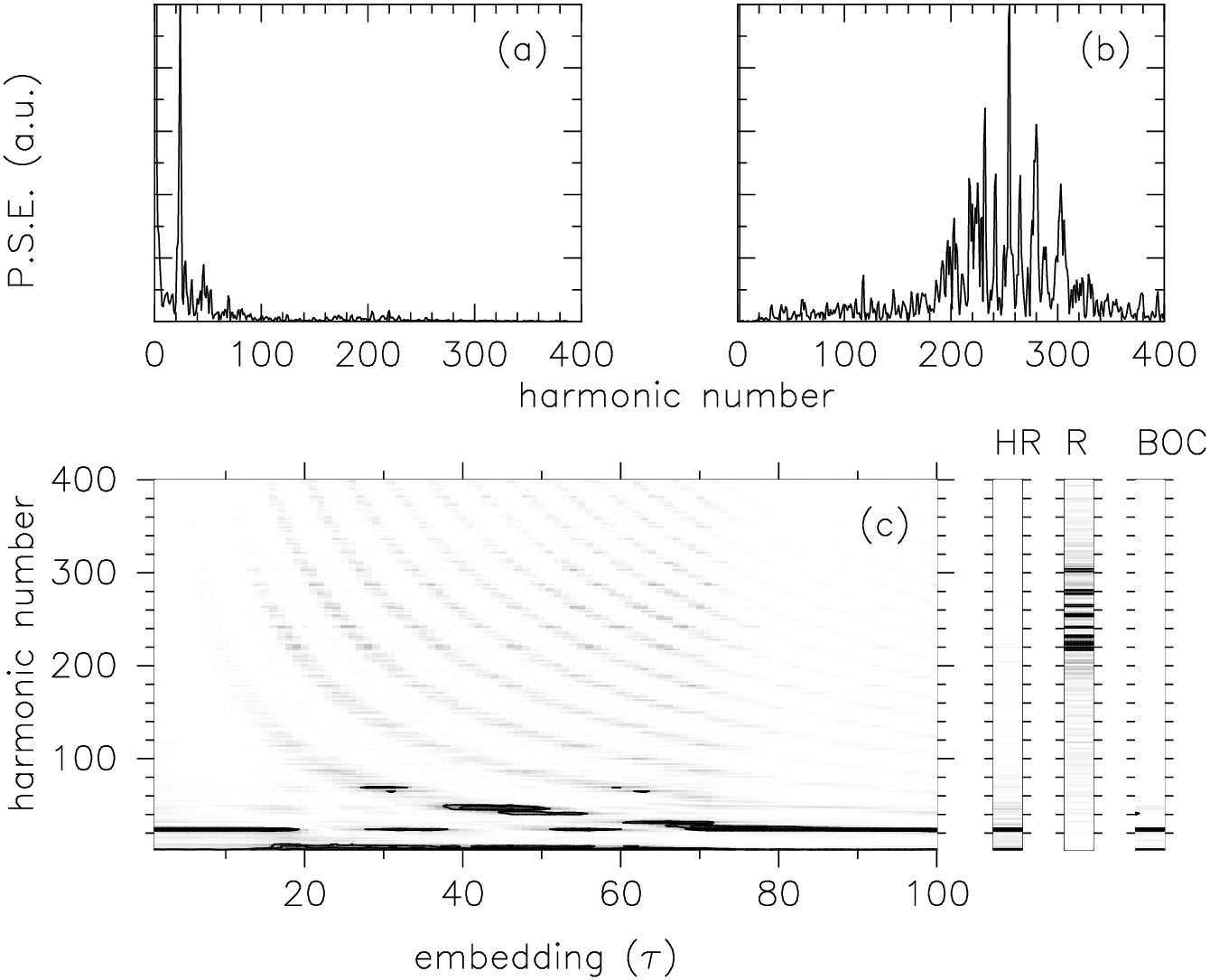,height=4.0in}
\caption{Spectral analysis and nonlinear coupling in the
cardiorrespiratory signals. Reconstruction using the R
signal.}
\end{center}
\end{figure}

\newpage

\begin{figure}[tbh]
\begin{center}
\
\psfig{file=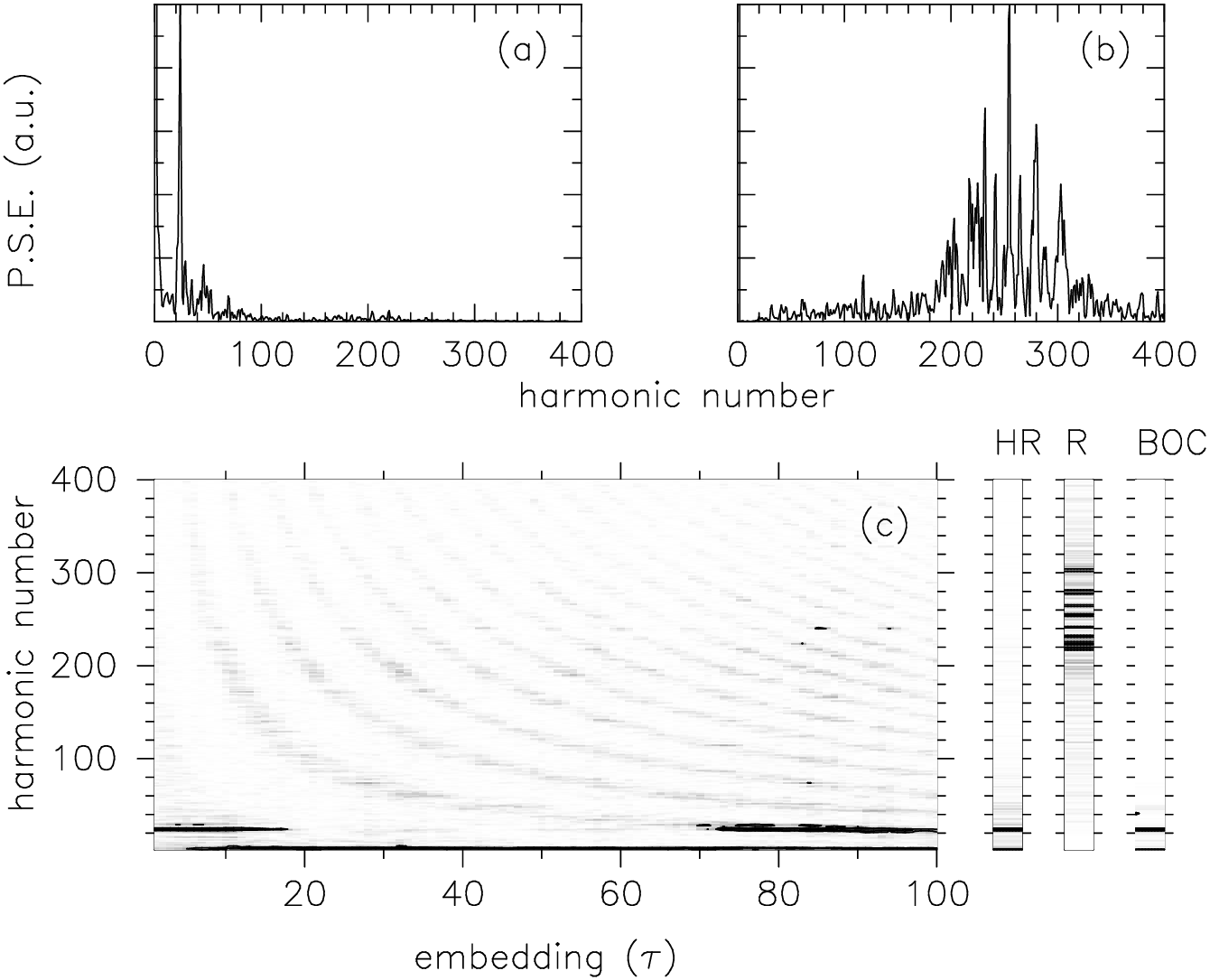,height=4.0in}
\caption{Spectral analysis and nonlinear coupling in the
cardiorrespiratory signals. Reconstruction using the HR
signal.}
\end{center}
\end{figure}

\end{document}